# Co-occurrence of Superparamagnetism and Anomalous Hall Effect in Highly Reduced Cobalt Doped Rutile TiO$_{2-\delta}$ Films

S. R. Shinde[1,*], S. B. Ogale[1,2], J. S. Higgins[1], H. Zheng[2], A. J. Millis[3], V. N. Kulkarni[1,+], R. Ramesh[1,2], R. L. Greene[1], and T. Venkatesan[1]

[1]Center for Superconductivity Research, Department of Physics, University of Maryland, College Park, MD 20742-4111
[2]Department of Materials and Nuclear Engineering, University of Maryland, College Park, MD 20742-42111
[3]Department of Physics, Columbia University, 538 West 120th Street, New York, New York 10027

We report a detailed magnetic and structural analysis of highly reduced Co doped rutile TiO$_{2-\delta}$ films displaying an anomalous Hall effect (AHE). The temperature and field dependence of magnetization, and transmission electron microscopy clearly establish the presence of nano-sized superparamagnetic cobalt clusters of ~8–10 nm size in the films at the interface. The co-occurrence of superparamagnetism and AHE raises questions regarding the use of the AHE as a test of the intrinsic nature of ferromagnetism in diluted magnetic semiconductors.

Oxide based diluted magnetic semiconductor (DMS) systems have recently attracted considerable attention because of the reports of room temperature ferromagnetism in several systems and their projected potential for spintronics devices [1–7]. Among these, cobalt doped TiO$_2$ in anatase and rutile forms has attracted particular interest, being the first oxide DMS system to be discovered [1]. However, subsequent reports have raised concerns about the initially suggested intrinsic nature of ferromagnetism in this material, due to the possibility of cobalt clustering under different growth conditions [3,8].

One of the important criteria for DMS to be intrinsic is suggested to be the observation of the anomalous Hall effect (AHE) in the material [9,10]. Only recently, different groups including ours, observed the AHE in highly reduced rutile TiO$_2$ films doped with either Fe or Co, raising optimism about the possibility of an intrinsic nature of ferromagnetism therein [11–13]. This situation clearly calls for detailed magnetic and microstructural studies of these films to establish or refute the premise of a strict connection between the AHE and the intrinsic nature of DMS ferromagnetism. Such work performed on our films and reported here clearly shows co-occurrence of superparamagnetism and the AHE. We emphasize that this work addresses only the highly reduced films in which the AHE is observed. It may be noted that to the best of our knowledge no AHE has yet been reported in anatase Ti$_{1-x}$Co$_x$O$_2$ films in which cobalt is suggested to be more uniformly distributed [1,2].

Rutile Ti$_{1-x}$Co$_x$O$_{2-\delta}$ films with x = 0, 0.01, 0.02, and 0.04 were grown by pulsed laser deposition on to R-Al$_2$O$_3$ substrates as described in Ref. [13]. In order to obtain high carrier concentration in the films to de-emphasize the ordinary Hall signal, the films were grown in high vacuum (base pressure ~ 2 x 10$^{-8}$ Torr). The results of AHE studies for Ti$_{0.98}$Co$_{0.02}$O$_{2-\delta}$ films have been previously reported elsewhere [13]. The magnetization (M) as a function of field (H) and temperature (T) was measured by superconducting quantum interference device (SQUID) magnetometry. The structural characterization of the films was performed with four circle x-ray diffraction (XRD), 1.8 MeV He$^+$ Rutherford backscattering spectrometry (RBS) and ion channeling, and transmission electron microscopy (TEM).

Figure 1(a) shows magnetic hysteresis loops measured at 5 and 300 K for a Ti$_{0.98}$Co$_{0.02}$O$_{2-\delta}$ film. It can be seen that although the hysteresis is observed at room temperature, the coercivity (H$_C$) and remanence (M$_R$) are very small. On the other hand, the film exhibits a large H$_C$ (> 230 Oe) and M$_R$ (~ 33 %) at 5 K. Figure 1(b) shows zero field cooled (ZFC) and field cooled (FC) M-T data for the same film measured at different fields. The FC and ZFC curves diverge substantially at low temperatures and the hump in ZFC curve progressively shifts to a lower temperature with increasing magnetic field. These behaviors are not expected for a ferromagnet and suggest the presence of magnetic nanoparticles in the films or the spin-glass nature of the system [14,15]. A large H$_C$ at 5 K [Fig. 1(a)] favors the presence of nanoparticles [16]. In nanoparticles, the formation of domain walls is energetically unfavorable and below a certain size, depending on the material, the particle stays in a single domain configuration. In single-domain particles, magnetization reversal can occur only by magnetization rotation as opposed to the combination of domain wall motion and magnetization rotation in a multi-domain particle. Since the magnetization rotation requires work to be done against the strong anisotropy forces, single-domain particle needs higher field to switch its magnetization and hence has a higher $H_C$ [17]. The magnetic hysteresis loops for the $Ti_{0.96}Co_{0.04}O_{2-\delta}$ films were similar to those of $Ti_{0.98}Co_{0.02}O_{2-\delta}$ films.

In spite of the fact that the magnetization measurements indicate the presence of magnetic nanoparticles in the film, we clearly observed the AHE for $Ti_{0.98}Co_{0.02}O_{2-\delta}$ films [13]. One such example is shown in the inset of Fig. 1(a). A sharp rise in the Hall resistivity at low field, i.e. AHE, is followed by a slow decrease corresponding to the ordinary Hall effect in n-type $TiO_2$.

The M-H data for films with lower (x = 0.01) cobalt concentration in the highly reduced films (Fig. 2) show further evidence for the presence of single domain particles as well as superparamagnetism. A large $H_C$ at 5 K, a rapid decrease of $H_C$ and $M_R$, and their disappearance above ~ 250 K [insets (b) and (c)] can be clearly noted. These features indicate the occurrence of superparamagnetism with a blocking temperature ($T_B$) of 250 K. The temperature dependence of M-H curves for $Ti_{0.98}Co_{0.02}O_{2-\delta}$ and $Ti_{0.96}Co_{0.04}O_{2-\delta}$ films is similar to that of $Ti_{0.99}Co_{0.01}O_{2-\delta}$ except with $T_B$ higher than 380 K [insets (b) and (c)], which could not be measured with SQUID. This sample also shows the AHE as shown in the inset (d).

In a single domain particle the easy directions of magnetization are separated by anisotropy barrier of magnitude $K_AV$, where $K_A$ is anisotropy energy density and V is particle volume. If the particle size is sufficiently small, above $T_B$ thermal fluctuations dominate and particle can spontaneously switch its magnetization from one easy axis to another. Such a system of superparamagnetic particles does not show hysteresis in the M-H curves above $T_B$; hence $H_C$ and $M_R$ are zero. Moreover, the magnetization curves measured above $T_B$ are expected to superimpose on each other when plotted as a function of H/T [18]. In the case of x = 0.01, the magnetization curves measured at T > 250 K clearly merge into each other [inset (a) of Fig. 2], confirming the occurrence of superparamagnetism above 250 K.

From the measured $T_B$, one can estimate the particle size by using the following relation [18,19]:
$$K_AV = 25k_BT_B$$
where, $k_B$ is Boltzmann constant. Assuming that the superparamagnetic particles are of metallic cobalt (confirmed by TEM as discussed later) and using $K_A$ = 4.5 x $10^6$ ergs/cm$^3$ [18], we estimate the particle diameter, D, to be ~ 7 nm corresponding to $T_B$ = 250 K for the $Ti_{0.99}Co_{0.01}O_{2-\delta}$ film. For $Ti_{0.98}Co_{0.02}O_{2-\delta}$ film, a higher $T_B$ suggests a particle size ~ 8 – 10 nm.

The TEM data for the $Ti_{0.98}Co_{0.02}O_{2-\delta}$ film (Fig. 3) enabled a direct observation of clustering and cluster size in the film. From electron diffraction pattern, these clusters were identified as cobalt metal clusters. Interestingly, these clusters are located at the film-substrate interface. The particles are about 9 – 10 nm in diameter and therefore are expected to show superparamagnetism.

All the results presented here confirm the presence of superparamagnetic cobalt nanoparticles in the highly reduced cobalt doped rutile $TiO_{2-\delta}$ films. The transport data on these films, reported elsewhere [13], suggest the presence of Magnéli phases in the films with a chemical formula $Ti_nO_{2n-1}$ (4 ≤ n ≤ 9) [20,21]. These phases are known to stabilize in the rutile structure with large concentration of oxygen vacancies, which might be responsible for cobalt diffusion and clustering during film growth.

In the light of the presented evidence on cobalt cluster formation, the observation of the AHE in the same films is rather surprising. One possibility to consider is that a small concentration of cobalt is distributed throughout the film, which causes the AHE signal. However, we did not find evidence of any significant quantity of uniformly dispersed cobalt from the RBS data. Fig. 3(c) shows the RBS spectrum (shown by symbols) for a $Ti_{0.98}Co_{0.02}O_{2-\delta}$ film. The solid line represents the simulated spectrum assuming that cobalt is uniformly distributed in the film. Inset of Fig. 3(c) explicitly shows that if cobalt were present in the top layers of the film, one should have seen it as indicated by the step in the simulation. Absence of the step indicates that cobalt is not present in the top layers of the film. This is consistent with the TEM results.

In the past observation of AHE has been reported in samples with magnetic clusters embedded in a nonmagnetic matrix [22]. Most such studies have been performed on samples with magnetic impurity concentration close to bulk percolation threshold whereas in our samples the average cobalt concentration is only ~ 2 %. Since in our samples cobalt clusters are mainly located at the interface, the average cluster density in the thin interface layer is higher, but even if one assumes that all of the 2 % cobalt atoms form ~ 10 nm clusters near the interface, one finds the percentage of occupied area to be ~ 7 % with intercluster (edge-edge) separation of ~ 20 nm. Thus we expect the Co clusters in our sample to be non-percolating. However, a simple argument (adapted from previous work of Shimshoni and Auerbach [23] regarding the observation of a quantized Hall effect in a non-percolating system) suggests that even in the non-percolating case an anomalous Hall effect will occur. Application of an electric field ($E_X$) along x-direction will lead in each magnetic nanoparticle to a y-direction current $J_y = \sigma_A M_Z E_X$ proportional to the z-component $M_Z$ of the particle magnetization. Here $\sigma_A$ is the anomalous Hall conductivity of the magnetic material (Co in our case). Since the surrounding $TiO_2$ matrix is conducting ($\rho_{300K} \sim 10^{-3}$ $\Omega$cm), if the droplet moments are aligned, each droplet would inject current in y-direction in to the $TiO_2$ matrix, and the current continuity equation then implies that this must be cancelled by development of a Hall voltage.

In conclusion, we have presented a clear evidence of superparamagnetism in highly reduced cobalt doped rutile films, which needs to be reconciled with the observation of the anomalous Hall effect (AHE) in the same films. This work questions the use of the AHE as an unambiguous test of the intrinsic nature of diluted magnetic semiconductor (DMS) without a detailed microscopic characterization of the sample. It is important to point out that the current work on highly reduced rutile cobalt doped $TiO_2$ is not a reflection on the possible intrinsic DMS character of cobalt doped $TiO_2$ (either anatase or rutile) grown under specific controlled conditions. To our knowledge, no AHE observation has yet been reported in such cases, however.

This work was supported under DARPA grant # N000140210962 and NSF-MRSEC grant # DMR 00-80008. The authors acknowledge the fruitful discussions with S. Das Sarma and S. Dhar (Maryland).

---------

[*] shinde@squid.umd.edu
[+] On leave from Indian Institute of Technology, Kanpur, India.

**Figure captions**

FIG. 1. (a) M-H curves for $Ti_{0.98}Co_{0.02}O_{2-\delta}$ film measured at 5 and 300 K. The inset shows Hall resistivity as a function of applied magnetic field; (b) field-cooled (FC, open symbols) and zero-field-cooled (ZFC, solid symbols) M-T curves measured at different fields.

FIG. 2. M-H curves for $Ti_{0.99}Co_{0.01}O_{2-\delta}$ film measured at different temperatures. The inset (a) shows magnetization as a function of H/T for different temperatures. In insets (b) and (c) are plotted the temperature dependence of $H_C$ and $M_R$ for $Ti_{0.99}Co_{0.01}O_{2-\delta}$ and $Ti_{0.98}Co_{0.02}O_{2-\delta}$ films. Inset (d) shows the magnetic field dependence of Hall resistivity of $Ti_{0.99}Co_{0.01}O_{2-\delta}$ film.

FIG. 3. TEM images [(a) and (b)] of $Ti_{0.98}Co_{0.02}O_{2-\delta}$ film at different magnifications. Some of the clusters are marked in image (b) by a black loop. (c) Experimentally observed (symbols) and simulated (line) RBS spectra for $Ti_{0.98}Co_{0.02}O_{2-\delta}$ film. The inset shows expanded view.

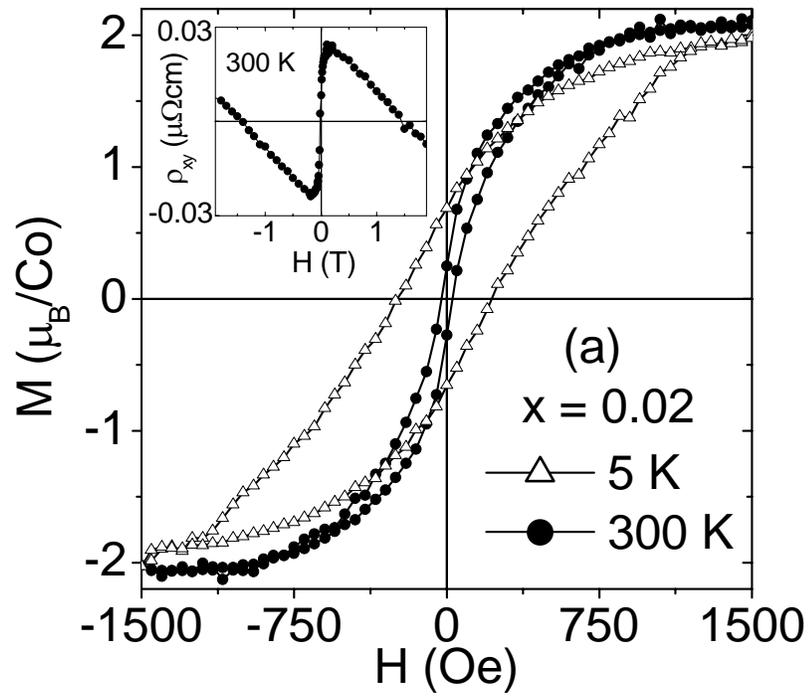

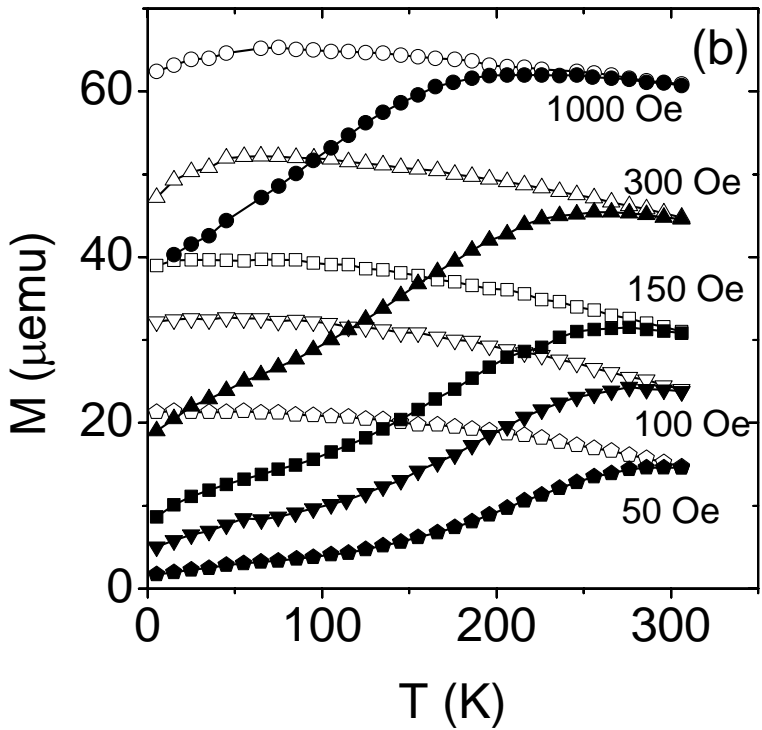

Figure 1, S.R. Shinde et al.

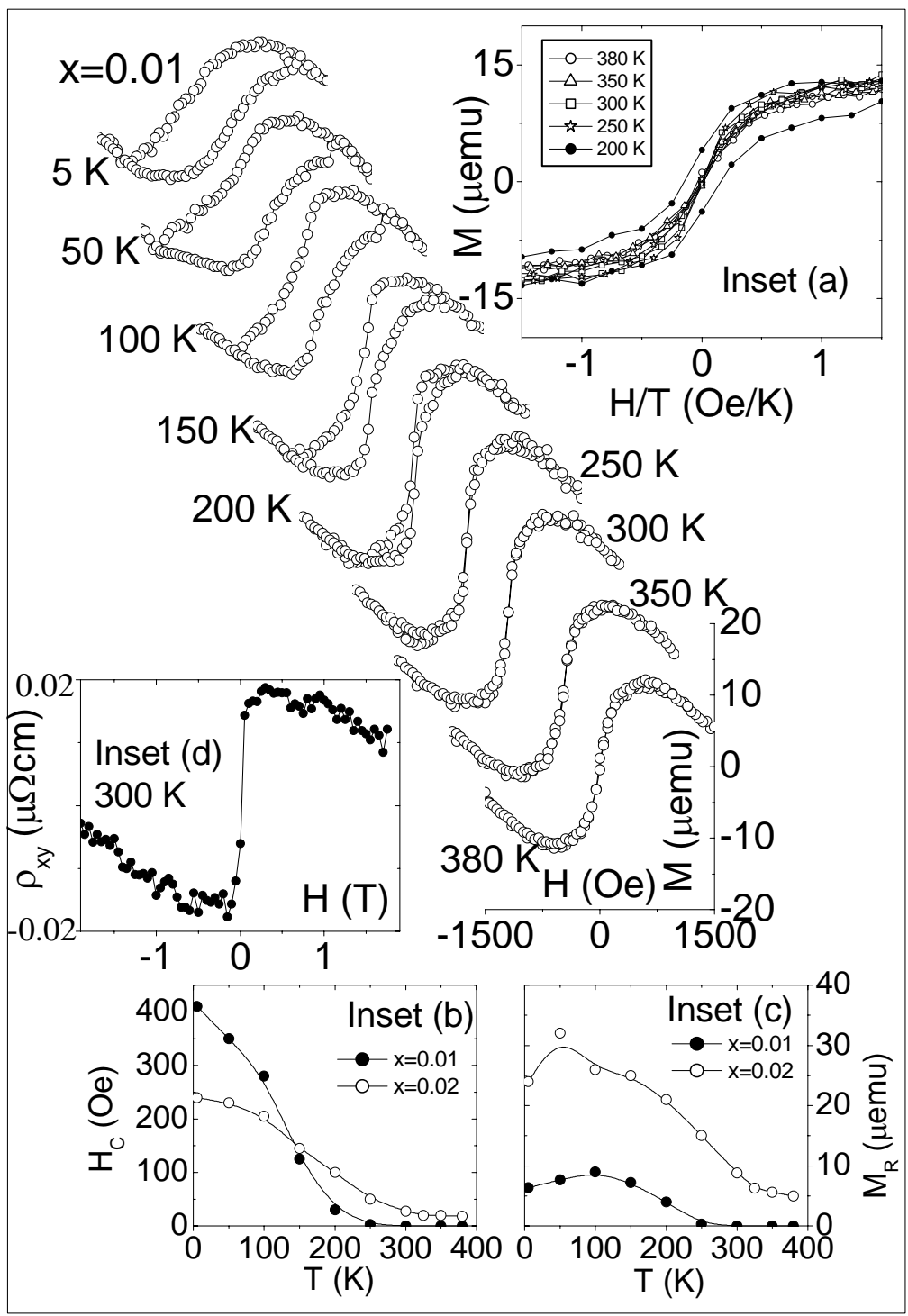

Figure 2, S. R. Shinde et al.

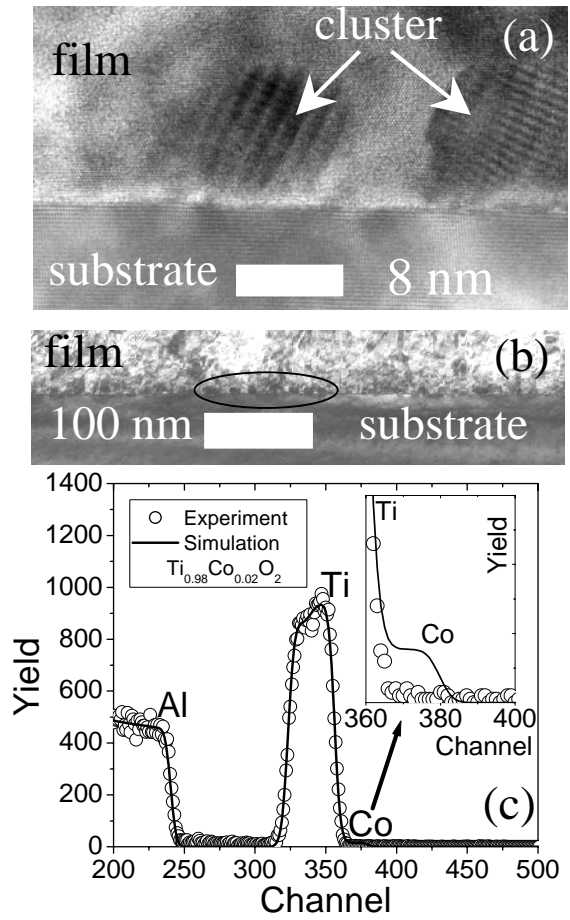

Figure 3, S. R. Shinde et al.